\documentclass[prd,aps,showpacs,nofootinbib]{revtex4}%
\usepackage{amsmath}
\usepackage{amsfonts}
\usepackage{amssymb}
\usepackage{bm}
\usepackage{graphicx}%
\begin{document}
\title{ Feynman propagator for the nonbirefringent CPT-even electrodynamics of the
Standard Model Extension}
\author{Rodolfo Casana$^{a}$, Manoel M. Ferreira Jr$^{a}$, Adalto R. Gomes$^{b}$,
Frederico E. P. dos Santos$^{a}$}
\affiliation{$^{a}$Departamento de F\'{\i}sica, Universidade Federal do Maranh\~{a}o
(UFMA), Campus Universit\'{a}rio do Bacanga, S\~{a}o Lu\'{\i}s - MA,
65085-580, Brazil.}
\affiliation{$^{b}$Departamento de F\'{\i}sica, Instituto Federal de Educa\c{c}\~{a}o,
Ci\^{e}ncia e Tecnologia do Maranh\~{a}o (IFMA), 65025-001, S\~{a}o Lu\'{\i}s,
Maranh\~{a}o, Brazil}

\begin{abstract}
The CPT-even gauge sector of the standard model extension is composed of
nineteen components comprised in the tensor $\left(  K_{F}\right)  _{\mu
\nu\rho\sigma}$, of \ which nine do not yield birefringence. In this work, we
examine the Maxwell electrodynamics supplemented by these nine nonbirefringent
CPT-even components in aspects related to the Feynman propagator and full
consistency (stability, causality, unitarity). We adopt a prescription that
parametrizes the nonbirefringent components in terms of a symmetric and
traceless tensor, $K_{\mu\nu},$ and second parametrization that writes
$K_{\mu\nu}$ in terms of two arbitrary four-vectors, $U_{\mu}$ and $V_{\nu}.$
We then explicitly evaluate the gauge propagator of this electrodynamics in a
tensor closed way. In the sequel, we show that this propagator and involved
dispersion relations can be specialized for the parity-odd\ and parity-even
sectors of the tensor $\left(  K_{F}\right)  _{\mu\nu\rho\sigma}$. \ In this
way, we reassess some results of the literature and derive some new outcomes
showing that the parity-even anisotropic sector engenders a stable, noncausal
and unitary electrodynamics.

\end{abstract}

\pacs{11.30.Cp, 12.60.-i, 11.55.Fv}
\maketitle

\section{Introduction}

Lorentz symmetry violation has been an issue of permanent interest in the past
few years, with many investigations in the context of the standard model
extension (SME) \cite{Colladay}, \cite{Samuel}. The SME incorporates terms of
Lorentz invariance violation (LIV) in all sectors of interaction and has been
studied in many respects \cite{General}, \cite{General2}. The investigations
in the context of the SME concern mainly the fermion sector
\cite{Fermion,Lehnert}, the gauge sector \cite{Jackiw}-\cite{Cherenkov2}, and
extensions involving gravity \cite{Gravity}. The violation of Lorentz symmetry
has also been addressed in other theoretical frameworks \cite{Jacobson}, with
many interesting developments \cite{Luca}, \cite{Other}.

The CPT-odd gauge sector of the SME, represented by the Carroll-Field-Jackiw
(CFJ) electrodynamics \cite{Jackiw}, has its properties largely examined in
literature, addressing consistency aspects \cite{Higgs} and modifications
induced in QED \cite{Adam,Soldati}, dimensional reduction \cite{Dreduction},
supersymmetry \cite{Susy}, controversies discussing the radiative generation
of the CFJ term \cite{Radio}, the vacuum emission of Cherenkov radiation
\cite{Cerenkov1}, the electromagnetic propagation in waveguides \cite{winder2}%
, modifications on the Casimir effect \cite{Casimir}, effects on the Planck
distribution and finite-temperature contributions \cite{winder1},
\cite{Tfinite}, possible effects on the anisotropies of the Cosmic Microwave
Background Radiation \cite{CMBR}. Since 2002 the CPT-even sector of SME has
been also much investigated, mainly in connection with issues able to provide
good bounds on its 19 LIV coefficients. The studies about the properties of
the CPT-even electrodynamics, represented by the tensor $\left(  K_{F}\right)
_{\alpha\nu\rho\varphi},$ were initiated by Kostelecky and Mewes in
Refs.\cite{KM1}, \cite{KM2}, where it was stipulated the existence of ten
linearly independent combinations of the components of $\left(  K_{F}\right)
_{\alpha\nu\rho\varphi}$ sensitive to birefringence. A broader and interesting
study in this respect was performed recently in Ref. \cite{Kostelec}. These
elements are contained in two $3\times3$ matrices, $\widetilde{\kappa}_{e+}$
and $\widetilde{\kappa}_{o-}.$ Using high-quality spectropolarimetry data of
cosmological sources \cite{KM3}, stringent upper bounds $\left(
10^{-32}\text{ and }10^{-37}\right)  $ were imposed on these birefringent LIV
parameters. Since 2003, precise experiments involving rotating optical and
microwave resonators have been performed \cite{Resonators}, yielding bounds at
the level of \ until one part in $10^{17}$ on the CPT-even parameters. The
study of Cherenkov radiation \cite{Cherenkov2} and the absence of emission of
Cherenkov radiation by ultrahigh energy cosmic rays \cite{Klink2,Klink3} has
been a point of great interest in latest years, as well as the photon-fermion
vertex interactions yielding new bounds on the LIV coefficients
\cite{Interac1,Interac2,Interac3}, \cite{Bocquet}. Investigations on
finite-temperature properties and the implied modifications on the Planck law
were developed as well for the CPT-even sector\cite{Tfinite2}, \cite{Tfinite3}%
. A full evaluation of the dispersion relations of the CPT-even
electrodynamics in connection with the birefringent role played by the LIV
coefficients is also presented in Refs. \cite{Tfinite2}, \cite{Tfinite3},
\cite{Prop}. More recently, the birefringence of the CPT-even coefficients at
leading and higher orders is discussed in Ref. \cite{Qasem}\textbf{.}

In a recent work, the gauge propagator of the CPT-even electrodynamics\ of the
SME has been\ explicitly carried out in the form of a $4\times4$ matrix
\cite{Prop}. The dispersion relations were determined from the poles of the
propagator, and used to analyze the stability, causality and unitarity of this
theory for the nonbirefringent parity-odd components and for the isotropic
parity-even one. \ The pole analysis showed that the parity-odd sector is
stable, noncausal, and unitary, whereas the parity-even isotropic sector,
represented uniquely by the trace component, provides a stable, causal, and
unitary theory for the range $0\leq\kappa_{\text{tr}}<1$.

In the present work, we evaluate the Feynman gauge propagator for the nine
nonbirefringent coefficients of the CTP-even sector of SME in an exact tensor
form, using the parametrization of the tensor $\left(  K_{F}\right)
_{\alpha\nu\rho\varphi}$ in terms of a symmetric and traceless rank-2 tensor,
$k_{\alpha\beta},$ introduced in Ref.\cite{Altschul}. In order to evaluate the
propagator, the tensor $k_{\alpha\beta}$, in turn, is parametrized in terms
of\ two four-vectors, $U_{\mu},V_{\nu},$ that contain the Lorentz-violating
components. Once the propagator is explicitly written, it is specialized for
the parity-odd and parity-even sectors of this theory for some choices of
$U_{\mu},V_{\nu},$ yielding the dispersion relations attained in
Ref.\cite{Prop}. New investigations concerning the anisotropic parity-even
components are performed, revealing that this sector is stable, noncausal, and unitary.

\section{Theoretical Model and the gauge propagator}

The Lagrangian of the CPT-even electrodynamics of SME is%
\begin{equation}
\mathrm{{\mathcal{L}}}=-\frac{1}{4}F_{\mu\nu}F^{\mu\nu}-\frac{1}{4}\left(
K_{F}\right)  _{\mu\nu\alpha\beta}F^{\mu\nu}F^{\alpha\beta}-\frac{1}{2\xi
}(\partial_{\mu}A^{\mu})^{2},\label{L1}%
\end{equation}
where $\xi$ is the gauge fixing parameter, $F_{\alpha\nu}$ is the
electromagnetic field tensor, and $\left(  K_{F}\right)  _{\alpha\nu
\rho\varphi}$ is a renormalizable and dimensionless coupling, responsible for
Lorentz violation. The tensor $\left(  K_{F}\right)  _{\alpha\nu\rho\varphi}$
has the same symmetries as the Riemann tensor $\left(  K_{F}\right)
_{\alpha\nu\rho\varphi}=-\left(  K_{F}\right)  _{\nu\alpha\rho\varphi},$
$\left(  K_{F}\right)  _{\alpha\nu\rho\varphi}=-\left(  K_{F}\right)
_{\alpha\nu\varphi\rho},~\left(  K_{F}\right)  _{\alpha\nu\rho\varphi}=\left(
K_{F}\right)  _{\rho\varphi\alpha\nu},$ $\left(  k_{F}\right)  _{\alpha
\beta\rho\varphi}+\left(  k_{F}\right)  _{\alpha\rho\varphi\beta}+\left(
k_{F}\right)  _{\alpha\varphi\beta\rho}=0,$ and a double null trace, $\left(
K_{F}\right)  ^{\rho\varphi}{}_{\rho\varphi}=0$. The tensor $\left(
K_{F}\right)  _{\alpha\nu\rho\varphi}$ has 19 independent components, from
which nine do not yield birefringence. A very useful parametrization for
addressing this theory is the one presented in Refs. \cite{KM1,KM2}, in which
these 19 components are contained in four $3\times3$ matrices:
\begin{align}
\left(  \widetilde{\kappa}_{e+}\right)  ^{jk} &  =\frac{1}{2}(\kappa
_{DE}+\kappa_{HB})^{jk},~~\left(  \widetilde{\kappa}_{e-}\right)  ^{jk}%
=\frac{1}{2}(\kappa_{DE}-\kappa_{HB})^{jk}-\frac{1}{3}\delta^{jk}(\kappa
_{DE})^{ii},~~\kappa_{\text{tr}}=\frac{1}{3}\text{tr}(\kappa_{DE}),~~\\
\left(  \widetilde{\kappa}_{o+}\right)  ^{jk} &  =\frac{1}{2}(\kappa
_{DB}+\kappa_{HE})^{jk},\text{ ~~}\left(  \widetilde{\kappa}_{o-}\right)
^{jk}=\frac{1}{2}(\kappa_{DB}-\kappa_{HE})^{jk},
\end{align}
where $\widetilde{\kappa}_{e}$ and $\widetilde{\kappa}_{o}$ designate
parity-even and parity-odd matrices, respectively. The $3\times3$ matrices
$\kappa_{DE},\kappa_{HB},\kappa_{DB},\kappa_{HE}$ are defined in terms of the
$\left(  K_{F}\right)  -$tensor components:
\begin{align}
\left(  \kappa_{DE}\right)  ^{jk} &  =-2\left(  K_{F}\right)  ^{0j0k},\text{
}\left(  \kappa_{HB}\right)  ^{jk}=\frac{1}{2}\epsilon^{jpq}\epsilon
^{klm}\left(  K_{F}\right)  ^{pqlm},\label{P1}\\
\text{ }\left(  \kappa_{DB}\right)  ^{jk} &  =-\left(  \kappa_{HE}\right)
^{kj}=\epsilon^{kpq}\left(  K_{F}\right)  ^{0jpq}.\label{P2}%
\end{align}
The matrices $\kappa_{DE},\kappa_{HB}$ contain together 11 independent
components while $\kappa_{DB},\kappa_{HE}$ possess together eight components,
which sums the 19 independent elements of the tensor $\left(  K_{F}\right)
_{\alpha\nu\rho\varphi}$. From these 19 coefficients, ten are sensitive to
birefringence and nine are nonbirefringent. These latter ones are contained in
the matrices $\widetilde{\kappa}_{o+}$ and $\widetilde{\kappa}_{e-}.$ The
analysis of birefringence data reveals the coefficients of the matrices
$\widetilde{\kappa}_{e+}$ and $\widetilde{\kappa}_{o-}$ are bounded to the
level of one part in $10^{32}$ \cite{KM1,KM2} and one part in $10^{37}%
$\cite{KM3}. This leads to the following constraints: $\kappa_{DE}%
=-\kappa_{HB},$ $\kappa_{DB}=\kappa_{HE},$ which implies%
\begin{equation}
\left(  \widetilde{\kappa}_{o+}\right)  ^{jk}=(\kappa_{DB})^{jk},\text{
}\left(  \widetilde{\kappa}_{e-}\right)  ^{jk}=(\kappa_{DE})^{jk}-\delta
^{jk}\kappa_{\text{tr}},\label{k_menos}%
\end{equation}
where the matrix $\widetilde{\kappa}_{e-}$ is symmetric and traceless (has
five components) and the matrix $\kappa_{DB}$ has now become antisymmetric,
possessing only three components written in terms of a 3-vector\cite{Kob},
\begin{equation}
\kappa^{i}=\frac{1}{2}\epsilon^{ipq}(\kappa_{DB})_{pq}.\label{kapa1}%
\end{equation}

An interesting way to parametrize the nine nonbirefringent components of the
tensor $(K_{F})$ is the one introduced in Ref. \cite{Altschul}, in which it is
written in terms of a symmetric traceless tensor $k_{\nu\rho}$:%
\begin{equation}
\left(  K_{F}\right)  ^{\lambda\nu\delta\rho}=\frac{1}{2}\left[
g^{\lambda\delta}k^{\nu\rho}-g^{\nu\delta}k^{\lambda\rho}+g^{\nu\rho
}k^{\lambda\delta}-g^{\lambda\rho}k^{\nu\delta}\right]  .\label{Alt}%
\end{equation}
Here, the nine nonbirefringent components are all contained in the symmetric
traceless tensor $k_{\nu\rho}$, defined as the contraction%
\begin{equation}
k^{\mu\nu}=\left(  K_{F}\right)  _{\alpha\text{ \ \ \ \ \ \ }}^{\text{ }%
\mu\alpha\nu}.
\end{equation}
The components of matrices $\kappa_{DE},\kappa_{HB},\kappa_{DB},\kappa_{HE}$
are linked with the components of the tensor $k^{\mu\nu}$ by means of the
following relations:
\begin{align}
\left(  \kappa_{DE}\right)  ^{jk} &  =\delta^{jk}k^{00}-k^{jk},\text{
\ }\left(  \kappa_{DB}\right)  ^{jk}=-\epsilon^{jkq}k^{0q},\text{ \ \ }\left(
\kappa_{HB}\right)  ^{jk}=-\delta^{jk}k^{ll}+k^{kj},\label{Mat1}\\
\left(  \widetilde{\kappa}_{o+}\right)  ^{jk} &  =-\epsilon^{jkq}k^{0q},\text{
}\left(  \widetilde{\kappa}_{e-}\right)  ^{jk}=\delta^{jk}k^{00}-k^{jk}%
-\delta^{jk}\kappa_{\text{tr}}.\label{Mat2}%
\end{align}
Considering that the $k^{\mu\nu}$ is traceless ($k_{\text{ }\mu}^{\mu}=0),$ it
holds $k^{00}=k^{ii}$, which leads to $\kappa_{DE}=-$ $\kappa_{HB}$ in
accordance with relation (\ref{Mat1}). Furthermore, the matrix $\kappa_{DB}$
is written in terms of the three components $k^{0q}.$ All this is consistent
with the nonbirefringent character of parametrization (\ref{Alt}). Regarding
the relations (\ref{kapa1}) and (\ref{Mat2}), it is possible to show that
\begin{equation}
k^{0q}=-\kappa^{q}.\label{kapa2}%
\end{equation}

The gauge propagator for Lagrangian density (\ref{L1}) was evaluated in a
matrix form in Ref. \cite{Prop} for the nonbirefringent parity-odd and the
isotropic parity-even component. To compute this gauge propagator in an exact
closed tensor form, we will use the prescription (\ref{Alt}). We begin writing
Lagrangian (\ref{L1}) in a squared form
\begin{equation}
\mathrm{{\mathcal{L}}}=~\frac{1}{2}A_{\mu}D^{\mu\nu}A_{\nu},
\end{equation}
where $D^{\mu\nu}=\square g^{\mu\nu}+\left(  1/\xi-1\right)  \partial^{\mu
}\partial^{\nu}-S^{\mu\nu}$ is a second order tensor operator, $S^{\mu\nu}$ is
the symmetric Lorentz-violating operator, $S^{\mu\nu}=2\left(  K_{F}\right)
^{\mu\alpha\beta\nu}\partial_{\alpha}\partial_{\beta},$ and $g^{\mu\nu
}=(+,---)$ is the metric tensor adopted here. The gauge propagator is the
defined as $\left\langle 0\left\vert T(A_{\mu}\left(  x\right)  A_{\nu}\left(
y\right)  )\right\vert 0\right\rangle =i\Delta_{\mu\nu}\left(  x-y\right)  ,$
where $\Delta_{\mu\nu}$ is the operator that fulfills the relation:
$D^{\mu\beta}\Delta_{\beta\nu}\left(  x-y\right)  =\delta^{\mu}{}_{\!\nu
}\delta\left(  x-y\right)  .$ We should compute the gauge propagator in the
Feynman gauge, $\xi=1$, which implies $D^{\mu\nu}=\square g^{\mu\nu}-S^{\mu
\nu}.$ Regarding the prescription (\ref{Alt}), the $S^{\lambda\rho}$ operator
becomes $S^{\lambda\rho}=\left[  g^{\lambda\delta}k^{\nu\rho}-g^{\nu\delta
}k^{\lambda\rho}+g^{\nu\rho}k^{\lambda\delta}-g^{\lambda\rho}k^{\nu\delta
}\right]  \partial_{\nu}\partial_{\delta}.$ In the Fourier representation, we
have $\widetilde{D}^{\lambda\rho}=-(p^{2}g^{\lambda\rho}+\tilde{S}%
^{\lambda\rho}),$ $\tilde{S}^{\lambda\rho}=-2\left(  K_{F}\right)
^{\lambda\nu\delta\rho}p_{\nu}p_{\delta},$ leading to%
\begin{align}
\tilde{S}^{\lambda\rho}  &  =-\left[  p^{\lambda}p_{\nu}k^{\nu\rho}%
-p^{2}k^{\lambda\rho}+p^{\rho}p_{\delta}k^{\lambda\delta}-g^{\lambda\rho
}p_{\delta}p_{\nu}k^{\nu\delta}\right]  ,\label{S2}\\
\widetilde{D}^{\lambda\rho}  &  =-p^{2}g^{\lambda\rho}+p^{\lambda}p_{\nu
}k^{\nu\rho}-p^{2}k^{\lambda\rho}+p^{\rho}p_{\delta}k^{\lambda\delta
}-g^{\lambda\rho}p_{\delta}p_{\nu}k^{\nu\delta}. \label{D1}%
\end{align}
For inverting this tensor operator, we must solve the relation $\tilde
{D}^{\lambda\rho}\tilde{\Delta}_{\rho\beta}=\delta^{\lambda}{}_{\!\beta}.$ For
it, we use the general parametrization for a symmetric traceless tensor,%
\begin{equation}
k^{\lambda\rho}=\frac{1}{2}(U^{\lambda}V^{\rho}+U^{\rho}V^{\lambda})-\frac
{1}{4}g^{\lambda\rho}(U\cdot V), \label{Symm}%
\end{equation}
in terms of two arbitrary four-vectors, $U^{\lambda},V^{\rho}$, which comprise
the Lorentz-violating coefficients. This prescription obviously assures the
traceless feature $\left(  k_{\text{ }\lambda}^{\lambda}=0\right)  ,$ as
expected. Moreover, it holds:
\begin{align}
k^{00}  &  =k^{ii}=\frac{3}{4}U^{0}V^{0}+\frac{1}{4}(\mathbf{U}\cdot
\mathbf{V}),\\
k^{ij}  &  =\frac{1}{2}(U^{i}V^{j}+U^{j}V^{j})+\frac{1}{4}\delta^{ij}%
(U^{0}V^{0}-\mathbf{U}\cdot\mathbf{V),}\\
\left(  \kappa_{DE}\right)  ^{jk}  &  =-\frac{1}{2}(U^{i}V^{j}+U^{j}%
V^{j})+\frac{1}{2}\delta^{ij}(U^{0}V^{0}+\mathbf{U}\cdot\mathbf{V).}
\label{kDE}%
\end{align}
Comparing Eq. (\ref{kDE}) with Eq. (\ref{k_menos}), we note that
\begin{equation}
\left(  \widetilde{\kappa}_{e-}\right)  ^{jk}=-\frac{1}{2}(U^{i}V^{j}%
+U^{j}V^{j}),\text{ \ \ }\kappa_{\text{tr}}=\frac{1}{2}(U^{0}V^{0}%
+\mathbf{U}\cdot\mathbf{V).} \label{kee}%
\end{equation}
Remembering that the matrix $\widetilde{\kappa}_{e-}$ is traceless, we should
impose $\mathbf{U}\cdot\mathbf{V}=0\mathbf{,}$ which simply implies
\begin{equation}
\kappa_{\text{tr}}=U^{0}V^{0}/2. \label{k_traco1}%
\end{equation}
After these preliminary definitions, we come back to the propagator
evaluation. Replacing the parametrization (\ref{Symm}) in Eq.(\ref{D1}), we
have:
\begin{align}
\widetilde{D}^{\lambda\rho}  &  =-\left[  p^{2}\left(  1-\frac{1}{2}U\cdot
V\right)  +(p\cdot U)(p\cdot V)\right]  g^{\lambda\rho}-\frac{1}{2}\left(
U\cdot V\right)  p^{\lambda}p^{\rho}+\frac{1}{2}(p\cdot U)\left(  p^{\rho
}V^{\lambda}+p^{\lambda}V^{\rho}\right) \nonumber\\
&  +\frac{1}{2}(p\cdot V)\left(  p^{\rho}U^{\lambda}+p^{\lambda}U^{\rho
}\right)  -\frac{1}{2}p^{2}\left(  U^{\lambda}V^{\rho}+U^{\rho}V^{\lambda
}\right)  .
\end{align}
In order to solve the relation $\tilde{D}^{\lambda\rho}\tilde{\Delta}%
_{\rho\beta}=\delta^{\lambda}{}_{\!\beta}$, we must first find a closed
operator algebra, composed by the following projectors:
\begin{equation}
\Theta_{\rho\beta},\text{ }\omega_{\rho\beta},\text{ }U_{\rho}V_{\beta},\text{
}U_{\beta}V_{\rho},\text{ }p_{\rho}U_{\beta},\text{ }p_{\beta}U_{\rho},\text{
}p_{\rho}V_{\beta},\text{ }p_{\beta}V_{\rho},V_{\beta}V_{\rho},U_{\beta
}U_{\rho},
\end{equation}
where $\Theta_{\mu\nu}=g_{\mu\nu}-\omega_{\mu\nu},\ \ \omega_{\mu\nu}=p_{\mu
}p_{\nu}/p^{2}$ are the transverse and longitudinal projectors. In this way,
it is proposed for the gauge propagator the general form:
\begin{equation}
\widetilde{\Delta}_{\rho\beta}\left(  p\right)  =(a_{1}\ \Theta_{\rho\beta
}+a_{2}\ \omega_{\rho\beta}+a_{3}U_{\rho}V_{\beta}+a_{4}U_{\beta}V_{\rho
}+a_{5}\ p_{\rho}U_{\beta}+a_{6}p_{\beta}U_{\rho}+a_{7}p_{\rho}V_{\beta}%
+a_{8}p_{\beta}V_{\rho}+a_{9}U_{\beta}U_{\rho}+a_{10}V_{\beta}V_{\rho}),
\end{equation}
with the coefficients $a_{i}$ being functions (of the momentum and of the
four-vectors $U_{\mu},V_{\nu}$) to be determined. The closed algebra of the
projectors is explicitly shown in Table I and Table II.

\begin{table}[ptb]
$%
\begin{tabular}
[c]{|c|c|c|c|c|c|}\hline
& $\ \ \ \ \ \ \ \ \ \ \ \ \ \ \ \ \ \Theta_{\rho\beta}$ & $\ \ \omega
_{\rho\beta}$ & $U_{\rho}V_{\beta}$ & $\ \ U_{\beta}V_{\rho}$ & $\ \ \ p_{\rho
}U_{\beta}$\\\hline
&  &  &  &  & \\\hline
$g^{\lambda\rho}$ & $\Theta_{\beta}^{\lambda}$ & $p^{\lambda}p_{\beta}/p^{2}$
& $U^{\lambda}V_{\beta}$ & $U_{\beta}V^{\lambda}$ & $\ \ p^{\lambda}U_{\beta}
$\\\hline
$\Theta^{\lambda\rho}$ & $\Theta_{\beta}^{\lambda}$ & $0$ & $U^{\lambda
}V_{\beta}-(p\cdot U)p^{\lambda}V_{\beta}/p^{2}$ & $U_{\beta}V^{\lambda
}-(p\cdot V)p^{\lambda}U_{\beta}/p^{2}$ & $0$\\\hline
$\omega^{\lambda\rho}$ & $0$ & $p^{\lambda}p_{\beta}/p^{2}$ & $(p\cdot
U)p^{\lambda}V_{\beta}/p^{2}$ & $(p\cdot V)p^{\lambda}U_{\beta}/p^{2}$ &
$p^{\lambda}U_{\beta}$\\\hline
$p^{\lambda}V^{\rho}$ & $p^{\lambda}V_{\beta}-\left(  V\cdot p\right)
p^{\lambda}p_{\beta}/p^{2}$ & $\left(  V\cdot p\right)  p^{\lambda}p_{\beta
}/p^{2}$ & $(U\cdot V)p^{\lambda}V_{\beta}$ & $V^{2}p^{\lambda}U_{\beta}$ &
$(p\cdot V)p^{\lambda}U_{\beta}$\\\hline
$p^{\rho}V^{\lambda}$ & $0$ & $V^{\lambda}p_{\beta}$ & $(p\cdot U)\ V_{\beta
}V^{\lambda}$ & $(p\cdot V)\ U_{\beta}V^{\lambda}$ & $p^{2}V^{\lambda}%
U_{\beta}$\\\hline
$p^{\lambda}U^{\rho}$ & $p^{\lambda}U_{\beta}-(p\cdot U)p^{\lambda}p_{\beta
}/p^{2}$ & $(p\cdot U)p^{\lambda}p_{\beta}/p^{2}$ & $U^{2}p^{\lambda}V_{\beta
}$ & $(U\cdot V)p^{\lambda}U_{\beta}$ & $(U\cdot p)p^{\lambda}U_{\beta}
$\\\hline
$p^{\rho}U^{\lambda}$ & $0$ & $U^{\lambda}p_{\beta}$ & $(p\cdot U)\ V_{\beta
}U^{\lambda}$ & $(p\cdot V)\ U_{\beta}U^{\lambda}$ & $p^{2}U^{\lambda}%
U_{\beta}$\\\hline
$p^{\lambda}p^{\rho}$ & $0$ & $\ p^{\lambda}p_{\beta}$ & $(p\cdot
U)p^{\lambda}V_{\beta}$ & $(p\cdot V)p^{\lambda}U_{\beta}$ & $p^{2}p^{\lambda
}U_{\beta}$\\\hline
$U^{\lambda}V^{\rho}$ & $\ U^{\lambda}V_{\beta}-(p\cdot V)U^{\lambda}p_{\beta
}/p^{2}$ & $(p\cdot V)U^{\lambda}p_{\beta}/p^{2}$ & $(U\cdot V)U^{\lambda
}V_{\beta}$ & $V^{2}U^{\lambda}U_{\beta}$ & $(p\cdot V)U^{\lambda}U_{\beta}%
$\\\hline
$U^{\rho}V^{\lambda}$ & $U_{\beta}V^{\lambda}-(p\cdot U)V^{\lambda}p_{\beta
}/p^{2}$ & $(p\cdot U)V^{\lambda}p_{\beta}/p^{2}$ & $U^{2}V^{\lambda}V_{\beta
}$ & $(U\cdot V)\ U_{\beta}V^{\lambda}$ & $(p\cdot U)V^{\lambda}U_{\beta}%
$\\\hline
\end{tabular}
$
\caption{Algebra of tensor projectors.}%
\label{table1}%
\end{table}

\begin{table}[ptb]
$%
\begin{tabular}
[c]{|c|c|c|c|c|c|}\hline
& $\ \ \ \ \ p_{\beta}U_{\rho}$ & $p_{\rho}V_{\beta}$ & $p_{\beta}V_{\rho}$ &
$U_{\beta}U_{\rho}$ & $V_{\beta}V_{\rho}$\\\hline
&  &  &  &  & \\\hline
$g^{\lambda\rho}$ & $\ \ \ \ \ p_{\beta}U^{\lambda}$ & $p^{\lambda}V_{\beta}$
& $p_{\beta}V^{\lambda}$ & $U_{\beta}U^{\lambda}$ & $V_{\beta}V^{\lambda}%
$\\\hline
$\Theta^{\lambda\rho}$ & $\ p_{\beta}U^{\lambda}-(p\cdot U)p^{\lambda}%
p_{\beta}/p^{2}$ & $0$ & $p_{\beta}V^{\lambda}-(p\cdot V)p^{\lambda}p_{\beta
}/p^{2}$ & $U_{\beta}U^{\lambda}-(p\cdot U)p^{\lambda}U_{\beta}/p^{2}$ &
$V_{\beta}V^{\lambda}-(p\cdot V)p^{\lambda}V_{\beta}/p^{2}$\\\hline
$\omega^{\lambda\rho}$ & $(p\cdot U)p^{\lambda}p_{\beta}/p^{2}$ & $p^{\lambda
}V_{\beta}$ & $(p\cdot V)p^{\lambda}p_{\beta}/p^{2}$ & $(p\cdot U)p^{\lambda
}U_{\beta}/p^{2}$ & $(p\cdot V)p^{\lambda}V_{\beta}/p^{2}$\\\hline
$p^{\lambda}V^{\rho}$ & $p^{\lambda}p_{\beta}(U\cdot V)$ & $p^{\lambda
}V_{\beta}(p\cdot V)$ & $p^{\lambda}p_{\beta}V^{2}$ & $(U\cdot V)p^{\lambda
}U_{\beta}$ & $V^{2}p^{\lambda}V_{\beta}$\\\hline
$p^{\rho}V^{\lambda}$ & $(p\cdot U)V^{\lambda}p_{\beta}$ & $p^{2}V^{\lambda
}V_{\beta}$ & $(p\cdot V)V^{\lambda}p_{\beta}$ & $(U\cdot p)V^{\lambda
}U_{\beta}$ & $(p\cdot V)V^{\lambda}V_{\beta}$\\\hline
$p^{\lambda}U^{\rho}$ & $U^{2}p^{\lambda}p_{\beta}$ & $(p\cdot U)p^{\lambda
}V_{\beta}$ & $(U\cdot V)p^{\lambda}p_{\beta}$ & $U^{2}p^{\lambda}U_{\beta}$ &
$(U\cdot V)p^{\lambda}V_{\beta}$\\\hline
$p^{\rho}U^{\lambda}$ & $(p\cdot U)U^{\lambda}p_{\beta}$ & $p^{2}U^{\lambda
}V_{\beta}$ & $(p\cdot V)U^{\lambda}p_{\beta}$ & $(p\cdot U)U_{\beta
}U^{\lambda}$ & $(p\cdot V)U^{\lambda}V_{\beta}$\\\hline
$p^{\lambda}p^{\rho}$ & $(p\cdot U)p^{\lambda}p_{\beta}$ & $p^{2}p^{\lambda
}V_{\beta}$ & $(p\cdot V)p^{\lambda}p_{\beta}$ & $(p\cdot U)p^{\lambda
}U_{\beta}$ & $(p\cdot V)p^{\lambda}V_{\beta}$\\\hline
$U^{\lambda}V^{\rho}$ & $(U\cdot V)U^{\lambda}p_{\beta}$ & $(p\cdot
V)U^{\lambda}V_{\beta}$ & $V^{2}U^{\lambda}p_{\beta}$ & $(U\cdot V)U_{\beta
}U^{\lambda}$ & $V^{2}U^{\lambda}V_{\beta}$\\\hline
$U^{\rho}V^{\lambda}$ & $U^{2}V^{\lambda}p_{\beta}$ & $(p\cdot U)V_{\beta
}V^{\lambda}$ & $(U\cdot V)p_{\beta}V^{\lambda}$ & $U^{2}V^{\lambda}U_{\beta}
$ & $(U\cdot V)V^{\lambda}V_{\beta}$\\\hline
\end{tabular}
$
\caption{Algebra of tensor projectors.}%
\label{Table2}%
\end{table}

\bigskip

Performing all the tensor contractions, we obtain a system of ten equations
for the ten coefficients $a_{i},$ whose solutions is%

\begin{equation}
a_{1}=-\frac{1}{[p^{2}-\frac{1}{2}p^{2}(U\cdot V)+(p\cdot U)(p\cdot
V)]},\text{ \ \ \ }a_{2}=a_{1}\left[  \frac{N}{\boxtimes(p)}\right]  ,
\end{equation}%
\begin{equation}
a_{3}=a_{4}=-\frac{a_{1}}{2}\left[  \frac{p^{2}[p^{2}+\frac{1}{2}(p\cdot
V)(p\cdot U)]}{\boxtimes(p)}\right]  ,
\end{equation}

\begin{equation}
a_{5}=a_{6}=\frac{a_{1}}{2}\left[  \frac{p^{2}(p\cdot V)+(p\cdot U)(p\cdot
V)^{2}-\frac{1}{2}(p\cdot U)V^{2}p^{2}}{\boxtimes(p)}\right]  ,
\end{equation}

\begin{equation}
a_{7}=a_{8}=\frac{a_{1}}{2}\left[  \frac{(p\cdot U)p^{2}+(p\cdot U)^{2}(p\cdot
V)-\frac{1}{2}U^{2}(p\cdot V)p^{2}}{\boxtimes(p)}\right]  ,
\end{equation}%
\begin{equation}
a_{9}=a_{10}=-\frac{a_{1}}{4}\left[  \frac{p^{2}[(p\cdot V)^{2}\ -p^{2}V^{2}%
]}{\boxtimes(p)}\right]  ,
\end{equation}
where the denominator element is
\begin{equation}
\boxtimes(p)=\text{\ }p^{4}(1-\frac{V^{2}U^{2}}{4})+\frac{p^{2}}{4}[4(p\cdot
U)(p\cdot V)+(p\cdot V)^{2}U^{2}+(p\cdot U)^{2}V^{2}]. \label{Caixa}%
\end{equation}
With these results, the gauge propagator is properly written as
\begin{align}
\left\langle 0\left\vert T(A_{\rho}\left(  x\right)  A_{\beta}\left(
y\right)  )\right\vert 0\right\rangle  &  =-\frac{i}{[p^{2}-\frac{1}{2}%
p^{2}(U\cdot V)+(p\cdot U)(p\cdot V)]\boxtimes(p)}\left\{  \frac{{}}{{}%
}\boxtimes(p)\Theta_{\rho\beta}+N(p)\omega_{\rho\beta}+F(p)(U_{\rho}V_{\beta
}+U_{\beta}V_{\rho})\right. \label{PropF}\\
&  \left.  \frac{{}}{{}}~\ \ \ \ \ \ \ \ \ \ \ \ \ \ +G(p)(p_{\rho}U_{\beta
}+p_{\beta}U_{\rho})+H(p)(p_{\rho}V_{\beta}+p_{\beta}V_{\rho})+I(p)U_{\beta
}U_{\rho}+L(p)V_{\beta}V_{\rho}\right\}  ,\nonumber
\end{align}
with the following coefficients:%
\begin{align}
G(p)  &  =\frac{1}{2}[(p\cdot V)p^{2}+(p\cdot U)(p\cdot V)^{2}-\frac{1}%
{2}(p\cdot U)p^{2}V^{2}],\\
H(p)  &  =\frac{1}{2}[(p\cdot U)p^{2}+(p\cdot V)(p\cdot U)^{2}-\frac{1}%
{2}(p\cdot V)p^{2}U^{2}],
\end{align}%
\begin{equation}
F(p)=-\frac{p^{2}}{2}[p^{2}+\frac{1}{2}(p\cdot V)(p\cdot U)],\text{
\ \ }I(p)=\frac{1}{4}p^{2}[p^{2}V^{2}-(p\cdot V)^{2}],\text{ \ \ }%
J(p)=\frac{1}{4}p^{2}[p^{2}U^{2}-(p\cdot U)^{2}], \label{FIL}%
\end{equation}%
\begin{align}
N(p)  &  =\frac{a_{1}}{4\boxtimes}\left[  4\boxtimes\lbrack1-\frac{1}%
{2}(U\cdot V)]-(p\cdot U)(p\cdot V)p^{2}U^{2}V^{2}+(p\cdot V)^{2}p^{2}%
U^{2}\right. \\
&  \left.  \frac{{}}{{}}~\ \ \ \ \ \ \ \ \ \ \ \ \ \ \ \ +\text{\ }(p\cdot
U)^{2}p^{2}V^{2}+(p\cdot U)(p\cdot V)^{3}U^{2}+(p\cdot V)(p\cdot U)^{3}%
V^{2}\right]  ,\nonumber
\end{align}
with $U^{2}=U\cdot U=U_{\mu}U^{\mu},V^{2}=V\cdot V=V_{\mu}V^{\mu}.$

Taking into account the expression (\ref{Symm}), an important comment is that
the products $U_{\beta}U_{\rho},V_{\beta}V_{\rho},U_{\rho}V_{\beta}$ are first
order terms in the Lorentz-violating coefficients of the matrix $k^{\lambda
\rho}.$ We thus notice that our exact results involve terms until third order
in the coefficients of the matrix $k^{\lambda\rho},$ although only second
order terms contribute to any observable associated with the matrix $S$. It is
still important to mention that this gauge propagator is symmetric before an
indices permutation $(\widetilde{\Delta}_{\rho\beta}=\widetilde{\Delta}%
_{\beta\rho})$ and before the $U\longleftrightarrow V$ permutation, as it
really must be.

\section{Dispersion relations}

The dispersion relations are read off from the poles of the propagator, that
is%
\begin{align}
\text{\ }p^{2}[1-\frac{1}{2}(U\cdot V)]+(p\cdot U)(p\cdot V) &  =0,\label{DR1}%
\\
p^{2}(1-\frac{V^{2}U^{2}}{4})+\frac{1}{4}[4(p\cdot U)(p\cdot V)+(p\cdot
V)^{2}U^{2}+(p\cdot U)^{2}V^{2}] &  =0.\label{DR2}%
\end{align}

From these relations we can analyze the energy stability, causality, and
unitarity of this theory. First, however, it is interesting to regard the
choices of $U^{\mu},$ $V^{\mu},$ that represent the parity-odd and parity-even
components. From now on, we adopt the general notation: $U^{\mu}%
=(U_{0},\mathbf{u}),$ $V^{\mu}=(V_{0},\mathbf{v}).$

We initiate discussing the isotropic parity-even coefficient, $\kappa
_{\text{tr}}$, that can be related only with the temporal components of
\ $U^{\mu},$ $V^{\mu}.$ Taking $U^{\mu}=(U_{0},0),$ $V^{\mu}=(V_{0},0)\ $as a
first choice, the tensor $k^{\lambda\rho}$ presents a single nonvanishing
component: $k^{00}=3U_{0}V_{0}/4=3\kappa_{\text{tr }}/2.$ The dispersion
relation (\ref{DR1}) yields
\begin{equation}
\text{\ }p_{0}=|\mathbf{p|}\sqrt{\frac{1-U_{0}V_{0}/2}{1+U_{0}V_{0}/2}},
\end{equation}
which has to be compared with the dispersion relation\ of Ref. \cite{Prop}
involving this isotropic component,\
\begin{equation}
\ p_{0}=|\mathbf{p|}\sqrt{(1-\kappa_{\text{tr}})/(1+\kappa_{\text{tr}})}.
\label{DR3B}%
\end{equation}
From this, we state the equality%
\begin{equation}
U_{0}V_{0}=2\kappa_{\text{tr }}, \label{A_B}%
\end{equation}
which coincides with Eq.(\ref{k_traco1}).\textbf{\ }The second dispersion
(\ref{DR2}) yields%
\begin{equation}
p_{0}^{2}=\mathbf{p}^{2}\frac{(4-U_{0}^{2}V_{0}^{2})}{4(1+U_{0}V_{0}%
)+U_{0}^{2}V_{0}^{2}}=\mathbf{p}^{2}\left[  \frac{2-U_{0}V_{0}}{2+U_{0}V_{0}%
}\right]  ,
\end{equation}
which is exactly reduced to Eq.(\ref{DR3B}) when the replacement (\ref{A_B})
is performed. This confirms the result of Ref. \cite{Prop}: Eq. (\ref{DR3B})
is the unique dispersion relation for the parity-even isotropic coefficient.

Taking now $U^{\mu}=(0,\mathbf{u}),$ $V^{\mu}=(V_{0},0),$ we specify the
parity-odd components, having $k^{0i}=\frac{1}{2}V_{0}u^{i}.$ In order to
verify it, we write the dispersion relation (\ref{DR1}) for this choice:
\begin{equation}
\text{\ \ }p_{0}^{2}=\mathbf{p}^{2}+p_{0}\mathbf{p\cdot(}V_{0}\mathbf{u}).
\end{equation}
This relation becomes equal to the dispersion relation of Ref. \cite{Prop} for
the parity-odd components represented in terms of \ the 3-vector
$\boldsymbol{\kappa,}$
\begin{equation}
p_{0}^{2}=\mathbf{p}^{2}-2p_{0}\left(  \mathbf{p}\cdot\boldsymbol{\kappa
}\right)  , \label{DR4}%
\end{equation}
whenever the following identification is done:
\begin{equation}
\boldsymbol{\kappa=-}\frac{1}{2}V_{0}\mathbf{u.} \label{Cond1}%
\end{equation}
It is easy to note that the relation (\ref{Cond1}) is consistent with Eq.
(\ref{kapa2}).

Into the choice $U^{\mu}=(0,\mathbf{u}),$ $V^{\mu}=(V_{0},0)$, the dispersion
relation (\ref{DR2}) is read as
\begin{equation}
4p_{0}^{2}-4p_{0}V_{0}(\mathbf{p\cdot u})=(4+V_{0}^{2}\mathbf{u}%
^{2})\mathbf{p}^{2}-(\mathbf{p\cdot u})^{2}V_{0}^{2}.\label{DR5}%
\end{equation}
Replacing the condition (\ref{Cond1}) in Eq. (\ref{DR2}), it turns out
\begin{equation}
p_{0}^{2}+2p_{0}(\mathbf{p\cdot\kappa})=\mathbf{p}^{2}+\mathbf{\kappa}%
^{2}\mathbf{p}^{2}-(\mathbf{p\cdot\kappa})^{2},\label{DR5A}%
\end{equation}
which is exactly the second dispersion relation for the parity-odd sector
attained in Ref.\cite{Prop}. Obviously, the parity-odd components can be also
particularized by $U^{\mu}=(U_{0},0),$ $\ V^{\mu}=(0,\mathbf{v}),$ for which
the dispersion relations (\ref{DR1}) and (\ref{DR2}) become
\begin{align}
\text{\ }p_{0}^{2} &  =\mathbf{p}^{2}+p_{0}\mathbf{p\cdot(}U_{0}%
\mathbf{v}),\label{DR6}\\
4p_{0}^{2}-4p_{0}U_{0}(\mathbf{p\cdot v}) &  =(4+U_{0}^{2}\mathbf{v}%
^{2})\mathbf{p}^{2}-(\mathbf{p\cdot v})^{2}U_{0}^{2}.\label{DR7}%
\end{align}
By replacing the condition $\boldsymbol{\kappa=-}\frac{1}{2}\left(
U_{0}\mathbf{v}\right)  $ in Eqs. (\ref{DR6}, \ref{DR7}), one recovers the
relations (\ref{DR4}, \ref{DR5A}) of Ref. \cite{Prop}. We thus notice that
both choices, [$U^{\mu}=(0,\mathbf{u}),$ $V^{\mu}=(V_{0},0)]$ or [$U^{\mu
}=(U_{0},0),$ $\ V^{\mu}=(0,\mathbf{v})],$ specify the parity-odd components
of the theory.

The third choice is the one that particularizes the anisotropic parity-even
components, $U^{\mu}=(0,\mathbf{u}),V^{\mu}=(0,\mathbf{v}).$ With it, the
dispersion relations (\ref{DR1}) and (\ref{DR2}) take the form
\begin{align}
p_{0}^{2} &  =\left[  \mathbf{p}^{2}-(\mathbf{p\cdot u})(\mathbf{p\cdot
v})\right]  ,\text{ \ \ \ }\label{DR8}\\
p_{0}^{2} &  =\mathbf{p}^{2}+\frac{1}{(4-\mathbf{u}^{2}\mathbf{v}^{2}%
)}[(\mathbf{p\cdot v)}^{2}\mathbf{u}^{2}+(\mathbf{p\cdot u})^{2}\mathbf{v}%
^{2}-4(\mathbf{p\cdot u})(\mathbf{p\cdot v})],\label{DR9}%
\end{align}
where the $\widetilde{\kappa}_{e-}$ traceless condition, $(\mathbf{u}%
\cdot\mathbf{v=0}),$ was taken into account. Such dispersion relations were
not evaluated in Ref.\cite{Prop}, once the anisotropic parity-even sector was
not analyzed there. However, these relations coincide with the ones of the
Appendix of Ref.\cite{Tfinite2} for $\left(  \mathbf{u}\cdot\mathbf{v}\right)
=0,$ except for a negative signal. In this reference, it was analyzed the
finite-temperature properties of this parity-even anisotropic electrodynamics
using the prescription $\left(  \widetilde{\kappa}_{e-}\right)  ^{jk}%
=(a^{j}b^{k}+a^{k}b^{j})/2,$ with $\mathbf{a\cdot b=0.}$ The relative signal
difference\textbf{\ }is compatible with Eq.(\ref{kee}). It may be recovered by
a suitable choice in which one of the vectors is taken as opposite, that is,
\ $\mathbf{u}\rightarrow-\ \mathbf{u}$, or $\mathbf{v}\rightarrow
-\ \mathbf{v}$ .

Thus, we can assert that the present prescription recovers all the exact
dispersion relations known for CPT-even electrodynamics, and states the new
relations (\ref{DR8}, \ref{DR9}).

\subsection{Causality and stability analysis}

\ In ref. \cite{Prop}, the dispersion relations (\ref{DR3B}, \ref{DR4}%
,\ref{DR5A}) were used to investigate the energy stability, causality and
unitarity of the CPT-even electrodynamics. It was verified that the parity-odd
sector represented by relations (\ref{DR4}, \ref{DR5A}) is stable, noncausal
and unitary, whereas the parity-even sector, described by relation
(\ref{DR3B}), is stable, causal and unitary for some limited values of
$\kappa_{\text{tr}}$. Once the dispersion relations here derived are shown to
recover the ones of Ref. \cite{Prop} for the parity-odd and parity-even
isotropic components, the consistency analysis performed for these two sectors
will not be retaken here. However, we now use the tensor propagator
(\ref{PropF}) for analyzing the dispersion relations and the consistency of
the parity-even anisotropic sector.

As it is known, the causality analysis is related to the sign of the
propagator poles \cite{Sexl}, given in terms of $\ p^{2},$ in such a way one
must have $p^{2}\geq0$ in order to preserve the causality (preventing the
existence of tachyons). We should now adopt a more detailed and confident
analysis on causality: the group velocity $(u_{g}=dp_{0}/d|\mathbf{p|})$\ and
the front velocity ($u_{front}=\lim_{|\mathbf{p|\rightarrow\infty}}u_{phase}%
)$.\ The causality is assured if $u_{g}\leq1$ and $u_{front}\leq1.$

In Ref. \cite{Prop}, the causality of the sector parity-odd was examined,
revealing a noncausal theory. The same kind of analysis showed that the
isotropic parity-even coefficient provides a causal theory for $0\leq
\kappa_{\text{tr}}<1.$

The\ causality of the anisotropic parity-even sector, however, was not
investigated, remaining to be verified. We take as starting point the
dispersion relations (\ref{DR8}) and (\ref{DR9}), which are now analyzed in
the following coordinate system:\ $x-$axis parallel to $\mathbf{u}$, $y-$axis
along $\mathbf{v,}$\ and the $z-$axis parallel to $\mathbf{u\times v}$. The
3-momentum expressed in spherical coordinates, $p=\left\vert \mathbf{p}%
\right\vert \left(  \sin\theta\cos\phi,\sin\theta\sin\phi,\cos\theta\right)
,$\ allows one to rewrite the dispersion relation (\ref{DR8}) as\textbf{\ }%
\begin{equation}
p_{0}=|\mathbf{p|}\sqrt{1-\frac{1}{2}\mathbf{|u|}|\mathbf{v}|\sin^{2}%
\theta\sin2\phi},
\end{equation}
which shows that the energy is always positive since the product
$\mathbf{|u|}|\mathbf{v}|$\ is small, so the stability is assured. The group
and front velocities are\textbf{\ }%
\begin{align}
u_{g} &  =\frac{dp_{0}}{d|\mathbf{p|}}=\sqrt{1-\frac{1}{2}\mathbf{|u|}%
|\mathbf{v}|\sin^{2}\theta\sin2\phi},\\
u_{front} &  =\sqrt{1-\frac{1}{2}\mathbf{|u|}|\mathbf{v}|\sin^{2}\theta
\sin2\phi},
\end{align}
\textbf{\ }Even for a small background ($\mathbf{|u|}|\mathbf{v}|<<1)$, for
$\phi\in\left\langle \pi/2,\pi\right\rangle \cup\left\langle 3\pi
/2,2\pi\right\rangle $ it may\ occur that $|u_{g}|>1$\ and $u_{front}>1$. So,
this model is in general noncausal.

For the relation (\ref{DR9}), we have%
\begin{equation}
p_{0}=|\mathbf{p|}\sqrt{1-\frac{\gamma}{2}\left[  \sin2\phi-\frac{1}%
{2}\mathbf{|u|}|\mathbf{v}|\right]  \mathbf{|u|}|\mathbf{v}|\sin^{2}\theta},
\end{equation}
where $\gamma=(1-\frac{1}{4}\mathbf{u}^{2}\mathbf{v}^{2})^{-1}.$ This relation
clearly indicates a positive energy for a small product $\mathbf{|u|}%
|\mathbf{v}|$, implying stability. The group and front velocities are given
by
\begin{equation}
u_{g}=u_{front}=\sqrt{1-\frac{\gamma}{2}\left[  \sin2\phi-\frac{1}%
{2}\mathbf{|u|}|\mathbf{v}|\right]  \mathbf{|u|}|\mathbf{v}|\sin^{2}\theta}.
\end{equation}
At the same way, this expression provides $|u_{g}|>1,u_{front}>1$\ for
$\phi\in\left\langle \pi/2,\pi\right\rangle \cup\left\langle 3\pi
/2,2\pi\right\rangle $. Thus, we conclude that the anisotropic parity-even
sector is stable but noncausal.

\subsection{Unitarity analysis}

The unitarity analysis of this model at tree-level is here carried out through
the saturation of the propagators with external currents \cite{Veltman}, which
must be implemented by means of the saturated propagator ($SP$), a scalar
quantity given as follows:
\begin{equation}
SP=J^{\ast\mu}\text{Res}(i\Delta_{\mu\nu})\text{ }J^{\nu}, \label{Sat2}%
\end{equation}
where Res$(i\Delta_{\mu\nu})$ is the matrix residue evaluated at the pole of
the propagator. The gauge current $(J^{\mu})$\ satisfies the conservation law
$\left(  \partial_{\mu}J^{\mu}=0\right)  ,$\ which in momentum space is read
as $p_{\mu}J^{\mu}=0$. In accordance with this method, the unitarity analysis
is assured whenever the imaginary part of the saturation $SP$\ (at the poles
of the propagator) is positive. A way to carry out the saturation consists in
determining the residue of the propagator matrix, evaluated at its own poles.

We begin writing the saturated gauge propagator (taking into account the
current conservation):%
\begin{equation}
SP=\left[  -iR\right]  \left[  \boxtimes(p)J^{2}+2F(p)\left(  U\cdot J\right)
\left(  V\cdot J\right)  +I(p)\left(  U\cdot J\right)  ^{2}+L(p)\left(  V\cdot
J\right)  ^{2}\right]  ,\label{Sat3}%
\end{equation}
where the terms $\boxtimes,F,I,L$, given by Eqs. (\ref{Caixa}- \ref{FIL}), are
to be evaluated in one of the poles of the propagator, and
\begin{equation}
R=\text{Res}\left[  \frac{1}{\left[  p^{2}(1-\displaystyle\frac{1}{2}U\cdot
V)+\left(  p\cdot{U}\right)  \left(  p\cdot V\right)  \right]  \boxtimes
}\right]  ,
\end{equation}
is the residue evaluated in the pole.

To examine the unitarity of the anisotropic parity-even sector, we use the
parametrization ${U}^{\mu}=(0,\mathbf{u}),$ ${V}^{\mu}=(0,\mathbf{v})$, where
$\mathbf{u}$ and $\mathbf{v}$ are orthogonal 3D-vectors,\ $\left(
\mathbf{u\cdot v}=0\right)  ,$ due to the traceless property of matrix
$\widetilde{\kappa}_{e-}$ [see Eq. (\ref{kee})].

We examine\ the pole stemming from the dispersion relation (\ref{DR1}),%
\begin{equation}
p^{2}=-\eta(p\cdot U)(p\cdot V),
\end{equation}
with $\eta=(1-\frac{1}{2}(U\cdot V))^{-1}.$ In the anisotropic sector,
$\eta=1.$ In this pole, the residue is
\begin{equation}
R=-\frac{1}{\frac{1}{4}p^{2}[(\mathbf{p}\cdot\mathbf{v})^{2}\mathbf{u}%
^{2}+(\mathbf{p}\cdot\mathbf{u})^{2}\mathbf{v}^{2}-\mathbf{u}^{2}%
\mathbf{v}^{2}(\mathbf{p}\cdot\mathbf{u})(\mathbf{p}\cdot\mathbf{v})]}.
\end{equation}
The saturation (\ref{Sat3}) is read as
\begin{equation}
SP=\left[  -iR\right]  \left[  R^{-1}J^{2}+2F\left(  U\cdot J\right)  \left(
V\cdot J\right)  +I\left(  U\cdot J\right)  ^{2}+L\left(  V\cdot J\right)
^{2}\right]  ,
\end{equation}
with%
\begin{align}
F &  =\frac{1}{4}p^{2}(\mathbf{p}\cdot\mathbf{v})(\mathbf{p}\cdot
\mathbf{u}),\\
\text{\ }I &  =\frac{1}{4}p^{2}(\mathbf{p}\cdot\mathbf{v})[(\mathbf{p}%
\cdot\mathbf{u})\mathbf{v}^{2}-(\mathbf{p}\cdot\mathbf{v})],\\
\text{\ }L &  =\frac{1}{4}p^{2}(\mathbf{p}\cdot\mathbf{u})[(\mathbf{p}%
\cdot\mathbf{v})\mathbf{u}^{2}-(\mathbf{p}\cdot\mathbf{u})].
\end{align}
Replacing all these expressions in the saturation, we achieve
\begin{equation}
SP=-\frac{iR}{4}\left\{  4R^{-1}J^{2}+(\mathbf{p}\cdot\mathbf{v}%
)(\mathbf{p}\cdot\mathbf{u})[\mathbf{u}^{2}(\mathbf{J}\cdot\mathbf{v}%
)^{2}+\mathbf{v}^{2}(\mathbf{J}\cdot\mathbf{u})^{2}]-[(\mathbf{p}%
\cdot\mathbf{v})(\mathbf{J}\cdot\mathbf{u})-(\mathbf{p}\cdot\mathbf{u}%
)(\mathbf{J}\cdot\mathbf{v})]^{2}\right\}  .\label{Sat4}%
\end{equation}

In order to verify the positivity of the expression above, it is suitable to
define a three-dimensional basis, generated by the vectors $\mathbf{\hat{v}}$,
$\mathbf{\hat{u}}$ and $\mathbf{\hat{c}}$:
\begin{equation}
\mathbf{\hat{v}=v/}|\mathbf{v}|~,~\ \mathbf{\hat{u}=u/|u|}~,\ \mathbf{\hat
{c}=(v\times u)/|u|}|\mathbf{v}|.
\end{equation}
In this basis, it holds
\begin{align}
\mathbf{J}\cdot\mathbf{v} &  \mathbf{=}J_{\text{v}}|\mathbf{v}|\text{;
\ }\mathbf{J}\cdot\mathbf{u=}J_{\text{u}}\mathbf{|u|}\text{; }\mathbf{p}%
\cdot\mathbf{v=}p_{\text{v}}|\mathbf{v}|\text{; }\mathbf{p}\cdot
\mathbf{u=}p_{\text{u}}\mathbf{|u|}\text{, }\label{J1}\\
J^{2} &  =J_{0}^{2}-\mathbf{J}^{2}=J_{0}^{2}-J_{\text{c}}^{2}-J_{\text{v}}%
^{2}-J_{\text{u}}^{2},\text{ }\label{JJ}\\
p_{0}^{2} &  =p_{\text{v}}^{2}+p_{\text{u}}^{2}+p_{\text{c}}^{2}%
-\mathbf{|u|}|\mathbf{v}|p_{\text{u}}p_{\text{v}}.\label{J3}%
\end{align}%
\begin{equation}
R=-\frac{1}{\frac{1}{4}\mathbf{|u|}^{2}|\mathbf{v}|^{2}[p_{\text{v}}%
^{2}+p_{\text{u}}^{2}-\mathbf{|u|}|\mathbf{v}|p_{\text{u}}p_{\text{v}}]}.
\end{equation}

From Eq.(\ref{Sat4}), and using the relations (\ref{J1}-\ref{J3}), we obtain:%
\begin{equation}
SP=-\frac{i}{4}R\left\{  4R^{-1}J^{2}+p_{\text{u}}p_{\text{v}}\mathbf{|u|}%
^{3}|\mathbf{v}|^{3}[J_{\text{v}}^{2}+J_{\text{u}}^{2}]-\mathbf{|u|}%
^{2}|\mathbf{v}|^{2}[p_{\text{v}}J_{\text{u}}-p_{\text{u}}J_{\text{v}}%
]^{2}\right\}  ,
\end{equation}
which is equivalent to:
\begin{equation}
SP=i\left\{  -J_{0}^{2}+J_{\text{c}}^{2}-\frac{\mathbf{|u|}^{2}|\mathbf{v}%
|^{2}}{4}R[p_{\text{v}}J_{\text{v}}+p_{\text{u}}J_{\text{u}}]^{2}\right\}  .
\end{equation}
Making more algebraic manipulations, and using the current conservation,
\ $p_{0}J_{0}=p_{\text{v}}J_{\text{v}}+p_{\text{u}}J_{\text{u}}+p_{\text{c}%
}J_{\text{c}},$ the saturation takes the form
\begin{equation}
SP=i\frac{\left[  J_{\text{c}}\left(  p_{\text{v}}^{2}+p_{\text{u}}%
^{2}-\mathbf{|u|}|\mathbf{v}|p_{\text{u}}p_{\text{v}}\right)  +p_{\text{c}%
}\left(  p_{\text{v}}J_{\text{v}}+p_{\text{u}}J_{\text{u}}\right)  \right]
^{2}}{\left[  p_{\text{v}}^{2}+p_{\text{u}}^{2}-\mathbf{|u|}|\mathbf{v}%
|p_{\text{u}}p_{\text{v}}\right]  \left[  p_{\text{v}}^{2}+p_{\text{u}}%
^{2}+p_{\text{c}}^{2}-\mathbf{|u|}|\mathbf{v}|p_{\text{u}}p_{\text{v}}\right]
}>0,
\end{equation}
which is compatible with the unitarity validity. In the result above, the
denominator term $p_{\text{v}}^{2}+p_{\text{u}}^{2}-\mathbf{|u|}%
|\mathbf{v}|p_{\text{u}}p_{\text{v}}$ was taken as positive. It occurs
whenever it holds the condition: $\mathbf{|u|}|\mathbf{v}|<2.$ As the
magnitude of the Lorentz-violating parameters is always much smaller than 1,
this condition is fulfilled. A similar development can be accomplished for the
pole stemming from the dispersion relation (\ref{DR2}). We thus assert that
the anisotropic parity-even sector is noncausal and unitary.

\section{Conclusions}

In this work, we have exactly evaluated the gauge propagator for the
nonbirefringent CPT-even electrodynamics of SME using a prescription proposed
in Ref. \cite{Altschul} and a parametrization for the symmetric $k^{\lambda
\delta}$ in terms of two arbitrary four-vectors. These parametrizations
allowed to obtain an exact tensor form for the propagator of the
nonbirefringent components which recovers the gauge propagator expressions is
the suitable parametrization choices are adopted. The involved dispersion
relations coincide with the ones obtained in Ref. \cite{Prop} for the
isotropic parity-even component and for the three parity-odd nonbirefringent
components. Furthermore, the dispersion relations for the anisotropic
parity-even components were achieved as well. \ The analysis of stability,
causality, and unitarity for the anisotropic parity-even components was
performed, revealing that this sector is stable, noncausal, and unitary. This
study completes the analysis initiated in Ref. \cite{Prop}. The achievement of
a tensor form propagator assures the facilities of the tensor calculus for
some interesting applications, as scattering amplitude evaluation in a Quantum
Electrodynamics context. Some investigations in this direction are now under development.

\begin{acknowledgments}
The authors are grateful for FAPEMA, CAPES and CNPq (Brazilian research
agencies) for invaluable financial support. The authors also thank Jose A.
Helayel-Neto for relevant comments on this work.
\end{acknowledgments}


\begin{thebibliography}{99}                                                                                               %


\bibitem {Colladay}D. Colladay and V. A. Kostelecky, Phys. Rev.\textit{ }D
\textbf{55}, 6760 (1997); D. Colladay and V. A. Kostelecky, Phys. Rev.\textit{
}D \textbf{58}, 116002 (1998); S. R. Coleman and S. L. Glashow, Phys. Rev.
D\textbf{\ 59}, 116008 (1999).

\bibitem {Samuel}V. A. Kostelecky and S. Samuel, Phys. Rev. Lett. \textbf{63},
224 (1989); Phys. Rev. Lett. \textbf{66}, 1811 (1991); Phys. Rev.\textit{
}D\textbf{\ 39}, 683 (1989); Phys. Rev.\textit{ }D\textbf{\ 40}, 1886 (1989),
V. A. Kostelecky and R. Potting, \textit{Nucl. Phys.} B\textbf{\ 359}, 545
(1991); Phys. Lett. B \textbf{381}, 89 (1996); V. A. Kostelecky and R.
Potting, Phys. Rev. D\ \textbf{51}, 3923 (1995).

\bibitem {General}N.M. Barraz, Jr., J.M. Fonseca, W.A. Moura-Melo, and J.A.
Helayel-Neto, Phys. Rev. D\textbf{76}, 027701 (2007); M. B. Cantcheff, Eur.
Phys. J. C \textbf{46}, 247 (2006); M. B. Cantcheff, C.F.L. Godinho, A.P.
Baeta Scarpelli, J.A. Helay\"{e}l-Neto, Phys. Rev. D \textbf{68}, 065025
(2003); H. Belich, T. Costa-Soares, J.A. Helayel-Neto, M.T.D. Orlando, R.C.
Paschoal, Phys. Lett. A \textbf{370}, 126 (2007); H. Belich, L.P. Colatto, T.
Costa-Soares, J.A. Helayel-Neto, M.T.D. Orlando, Eur. Phys. J. C \textbf{62},
425 (2009); C. N. Ferreira, J. A. Helay\"{e}l-Neto and C. E. C. Lima, New J.
Phys. \textbf{12} (2010) 053029.

\bibitem {General2}V.A. Kostelecky, N. Russell, J. Tasson,
Phys.Rev.Lett.\textbf{100}, 111102 (2008); V.A. Kostelecky, N. Russell,
Phys.Lett.B \textbf{693}, 443 (2010); B. Altschul, Q.G. Bailey, V.A.
Kostelecky, Phys.Rev.D \textbf{81}, 065028 (2010); B. Altschul, Phys.Rev.D
\textbf{82}, 016002 (2010); S. Herrmann, A. Senger, K. Mohle, M. Nagel, E.V.
Kovalchuk, A. Peters, Phys.Rev.D\textbf{ 80},105011 (2009); A. Ferrero, B.
Altschul, Phys.Rev.D \textbf{80},125010 (2009); Z. Xiao, B-Q. Ma, Phys.Rev.D
\textbf{80}, 116005 (2009); \ K-Y. Chung, S.-w. Chiow, S. Herrmann, S. Chu, H.
Muller, Phys.Rev.D \textbf{80}, 016002 (2009).

\bibitem {Fermion}B. Altschul, Phys. Rev. D \textbf{70}, 056005 (2004); G. M.
Shore, Nucl. Phys. B \textbf{717}, 86 (2005); \ D. Colladay and V. A.
Kostelecky, Phys. Lett. B \textbf{511}, 209 (2001); O. G. Kharlanov and V. Ch.
Zhukovsky, J. Math. Phys. \textbf{48}, 092302 (2007); R. Lehnert, Phys. Rev. D
\textbf{68}, 085003 (2003); V.A. Kostelecky and C. D. Lane, J. Math. Phys.
\textbf{40}, 6245 (1999); R. Lehnert, J. Math. Phys. \textbf{45}, 3399 (2004);
W. F. Chen and G. Kunstatter, Phys. Rev. D \textbf{62}, 105029 (2000); B.
Goncalves, Y. N. Obukhov, I. L. Shapiro, Phys.Rev.D \textbf{80}, 125034 (2009).

\bibitem {Lehnert}V. A. Kostelecky and R. Lehnert, Phys. Rev. D\textbf{\ 63,
}065008 (2001)\textbf{.}

\bibitem {Jackiw}S.M. Carroll, G.B. Field and R. Jackiw, Phys. Rev.\textit{\ }%
D \textbf{41}, 1231 (1990).

\bibitem {Higgs}A. P. Baeta Scarpelli, H. Belich, J. L. Boldo, J.A.
Helayel-Neto, Phys. Rev. D\textit{\ }\textbf{67}, 085021 (2003); A.P. Baeta
Scarpelli and J.A. Helayel-Neto, Phys.Rev. D \textbf{73}, 105020 (2006).

\bibitem {Adam}C. Adam and F. R. Klinkhamer, Nucl. Phys.\textit{\ }B
\textbf{607}, 247 (2001); C. Adam and F. R. Klinkhamer, Nucl. Phys.\textit{\ }%
B \textbf{657}, 214 (2003).

\bibitem {Soldati}A.A. Andrianov and R. Soldati, Phys. Rev. D \textbf{51},
5961 (1995); Phys. Lett. B \textbf{435}, 449 (1998); A.A. Andrianov, R.
Soldati and L. Sorbo, Phys. Rev. D \textbf{59}, 025002 (1998); A. A.
Andrianov, D. Espriu, P. Giacconi, R. Soldati, J. High Energy Phys.
\textbf{0909}, 057 (2009); J. Alfaro, A.A. Andrianov, M. Cambiaso, P.
Giacconi, R. Soldati, Int.J.Mod.Phys.A \textbf{25}, 3271 (2010); V. Ch.
Zhukovsky, A. E. Lobanov, E. M. Murchikova, Phys.Rev. D\textbf{73} 065016, (2006).

\bibitem {Dreduction}H. Belich, M.M. Ferreira Jr., J.A. Helayel-Neto, M.T.D.
Orlando, Phys. Rev. D \textbf{67},125011 (2003); Erratum-ibid., Phys. Rev.
\textbf{D }69, 109903 (2004).

\bibitem {Susy}M.S. Berger, V. A. Kostelecky, Phys.Rev.D \textbf{65}, 091701
(2002); H. Belich , J.L. Boldo, L.P. Colatto, J.A. Helayel-Neto, A.L.M.A.
Nogueira, Phys.Rev. D \textbf{68}, 065030 (2003); A.P. Baeta Scarpelli, H.
Belich, J.L. Boldo, L.P. Colatto, J.A. Helayel-Neto, A.L.M.A. Nogueira, Nucl.
Phys. Proc. Suppl. \textbf{127}, 105 (2004).

\bibitem {Radio}R. Jackiw and V. A. Kosteleck\'{y}, Phys. Rev. Lett.
\textbf{82}, 3572 (1999); J. M. Chung and B. K. Chung Phys. Rev.
D\textbf{\ 63}, 105015 (2001); J.M. Chung, Phys.Rev. D\ \textbf{60}, 127901
(1999); G. Bonneau, Nucl.Phys. B\ \textbf{593}, 398 (2001); M. Perez-Victoria,
Phys. Rev. Lett. \textbf{83}, 2518 (1999); M. Perez-Victoria, J. High. Energy
Phys. \textbf{\ 0104}, (2001) 032; O.A. Battistel and G. Dallabona, Nucl.
Phys. B \textbf{610}, 316 (2001); O.A. Battistel and G. Dallabona, J. Phys. G
\textbf{28}, L23 (2002); J. Phys. G \textbf{27}, L53 (2001); A. P. B.
Scarpelli, M. Sampaio, M. C. Nemes, and B. Hiller, Phys. Rev. D\ \textbf{64},
046013 (2001); T. Mariz, J.R. Nascimento, E. Passos, R.F. Ribeiro and F.A.
Brito, J. High. Energy Phys. \textbf{0510} (2005) 019; J. R. Nascimento, E.
Passos, A. Yu. Petrov, F. A. Brito, J. High. Energy Phys. \textbf{0706},
(2007) 016; B. Altschul, Phys. Rev. D \textbf{70}, 101701 (2004); A.P.B.
Scarpelli, M. Sampaio, M.C. Nemes, B. Hiller, Eur. Phys. J. C \textbf{56}, 571
(2008); \ F.A. Brito, J.R. Nascimento, E. Passos, A.Yu. Petrov, Phys. Lett.
\textbf{B} 664, 112 (2008); Oswaldo M. Del Cima, J. M. Fonseca, D.H.T. Franco,
O. Piguet, Phys. Lett. \textbf{B} 688, 258 (2010); F.A. Brito, L.S. Grigorio,
M.S. Guimaraes, E. Passos, C. Wotzasek, Phys.Rev. D \textbf{78}, 125023 (2008).

\bibitem {Cerenkov1}R. Lehnert and R. Potting, Phys. Rev. Lett. \textbf{93},
110402 (2004); R. Lehnert and R. Potting, Phys. Rev. D \textbf{70}, 125010
(2004); B. Altschul, Phys. Rev. D \textbf{75}, 105003 (2007); C. Kaufhold and
F.R. Klinkhamer, Nucl. Phys. B \textbf{734, 1 }(2006).

\bibitem {winder2}A.H. Gomes, J.M. Fonseca, W.A. Moura-Melo, A.R. Pereira,
JHEP \textbf{05}, 104 (2010).

\bibitem {Casimir}M. Frank and I. Turan, Phys. Rev. D \textbf{74}, 033016
(2006); O.G. Kharlanov, V.Ch. Zhukovsky, Phys. Rev. D \textbf{81}, 025015 (2010).

\bibitem {winder1}J. M. Fonseca, A. H. Gomes, W. A. Moura-Melo, Phys. Lett. B
\textbf{671}, 280 (2009).

\bibitem {Tfinite}R. Casana, M. M. Ferreira Jr. and J. S. Rodrigues, Phys.
Rev. D \textbf{78}, 125013 (2008); M. Gomes, T. Mariz, J. R. Nascimento, A.Yu.
Petrov, A. F. Santos, and A. J. da Silva, Phys. Rev. D \textbf{81}, 045013
(2010); F.A. Brito, L.S. Grigorio, M.S. Guimaraes , E. Passos, C. Wotzasek,
Phys. Lett. B \textbf{681}, 495(2009).

\bibitem {CMBR}V.A. Kostelecky, M.Mewes, \ Phys.Rev.Lett. \textbf{99}, 011601
(2007); V. A. Kostelecky and M. Mewes, Astrophys. J. Lett. \textbf{689}, L1
(2008); J.-Q. Xia, Hong Li, X. Wang, X. Zhang, Astron.Astrophys. \textbf{483},
715 (2008); J.-Q. Xia, H. Li, X. Zhang, Phys.Lett. B \textbf{687}, 129(2010);
B. Feng, M. Li, J.-Q. Xia, X. Chen, X. Zhang, Phys. Rev. Lett. \textbf{96},
221302 (2006); P. Cabella, P. Natoli, J. Silk, Phys. Rev. D \textbf{76},
123014 (2007).

\bibitem {KM1}V. A. Kostelecky and M. Mewes, Phys. Rev. Lett. \textbf{87},
251304 (2001).

\bibitem {KM2}V. A. Kostelecky and M. Mewes, Phys. Rev. D\textbf{\ 66}, 056005 (2002).

\bibitem {Kostelec}V. A. Kostelecky and M. Mewes, Phys. Rev.\textit{\ }D
\textbf{80}, 015020\ (2009).

\bibitem {KM3}V. A. Kostelecky and M. Mewes, Phys. Rev. Lett. \textbf{97},
140401 (2006).

\bibitem {Resonators}H.Muller, S. Herrmann, C.Braxmaier, S. Schiller, A.
Peters, Phys. Rev. Lett. \textbf{91,} 020401 (2003); H. M\"{u}ller, S.
Herrmann, A. Saenz, A. Peters, and C. L\"{a}mmerzahl, Phys. Rev. D
\textbf{68}, 116006 (2003); H. Muller, C. Braxmaier, S. Herrmann, A. Peters,
and C.L\"{a}mmerzahl, Phys. Rev. D\textbf{ 67}, 056006 (2003); J. A. Lipa, S.
Wang, D. A. Stricker, D. Avaloff, Phys.Rev. Lett. \textbf{90}, 060403 (2003);
H. M\"{u}ller, Phys. Rev. D \textbf{71}, 045004 (2005); S. Herrmann, A.
Senger, E. Kovalchuk, H. M\"{u}ller, A. Peters, Phys. Rev. Lett. \textbf{95},
150401 (2005); P. Wolf, M. E. Tobar, S. Bize, A. Clairon, A. N. Luiten, G.
Santarelli, Phys. Rev. D \textbf{70}, 051902 (2004); P. L. Stanwix, M. E.
Tobar, P. Wolf, M. Susli, C. R. Locke, E. N. Ivanov, J. Winterflood, F. van
Kann, Phys. Rev. Lett. \textbf{95}, 040404 (2005); P.L. Stanwix, M. E. Tobar,
P. Wolf, C. R. Locke, E. N. Ivanov, Phys. Rev. D \textbf{74}, 081101 (R)
(2006); H. Muller, P. L. Stanwix, M. E. Tobar, E. Ivanov, P. Wolf, S.
Herrmann, A. Senger, E. Kovalchuk, A. Peters, Phys.Rev.Lett. \textbf{99},
050401 (2007); S. Herrmann, A. Senger, K. M\"{o}hle, M. Nagel, E. Kovalchuk,
A. Peters, Phys. Rev. D \textbf{80}, 105011 (2009); Ch. Eisele, A.Yu. Nevsky,
and S. Schiller, Phys. Rev. Lett. \textbf{103}, 090401 (2009); Q. Exirifard,
"Triangular Fabry-Perot resonator", arxiv:1010.2057.

\bibitem {Masers}D. F. Phillips, M. A. Humphrey, E. M. Mattison, R. E. Stoner,
R. F.C. Vessot, and R. L. Walsworth, Phys. Rev. D 63, 111101 (2001); D. Bear,
R. E. Stoner, R. L. Walsworth, V. A. Kostelecky, and C. D. Lane, Phys. Rev.
Lett. \textbf{85}, 5038 (2000); 89, 209902(E) (2002); M. A. Humphrey, D. F.
Phillips, E. M. Mattison, R. F. C. Vessot, R. E. Stoner, and R. L.Walsworth,
Phys. Rev. A \textbf{68}, 063807 (2003).

\bibitem {Cherenkov2}B. Altschul, Nucl. Phys. B\textbf{\ 796}, 262 (2008); B.
Altschul, Phys. Rev. Lett. \textbf{98}, 041603 (2007); C. Kaufhold and F.R.
Klinkhamer, Phys. Rev. D \textbf{76}, 025024 (2007)\textbf{. }

\bibitem {Gravity}V.Alan Kostelecky, Phys.Rev.D \textbf{69}, 105009 (2004); V.
A. Kosteleck\'{y}, Neil Russell, and Jay D. Tasson, Phys. Rev. Lett.
\textbf{100}, 111102 (2008); V. A. Kosteleck\'{y} and J. D. Tasson, Phys. Rev.
Lett. \textbf{102}, 010402 (2009); Q. G. Bailey, V.A. Kostelecky, Phys.Rev. D
\textbf{74}, 045001 (2006); Q. G. Bailey, Phys.Rev. D \textbf{80}, 044004
(2009); V.A. Kostelecky and R. Potting, Phys.Rev. D\textbf{79}, 065018 (2009);
Q. G. Bailey, Phys. Rev. D \textbf{82}, 065012 (2010); V.B. Bezerra, C.N.
Ferreira, J.A. Helayel-Neto, Phys.Rev. D \textbf{71}, 044018 (2005); J.L.
Boldo, J.A. Helayel-Neto, L.M. de Moraes, C.A.G. Sasaki, V.J. Vasquez Otoya,
Phys. Lett. B \textbf{689}, 112 (2010); A.F. Ferrari, M. Gomes, J.R.
Nascimento, E. Passos, A.Yu. Petrov, A. J. da Silva, Phys. Lett. B
\textbf{652}, 174 (2007).

\bibitem {Jacobson}T. Jacobson, S. Liberati, D. Mattingly, Annals Phys.
\textbf{321}, 150 (2006); T. Jacobson, S. Liberati, D. Mattingly, Phys.Rev.
D\textbf{67}, 124011 (2003); T. Jacobson, S. Liberati, D. Mattingly, Nature
\textbf{424}, 1019 (2003); T. Jacobson, S. Liberati, D. Mattingly, Phys.Rev. D
\textbf{67}, 124011 (2003).

\bibitem {Luca}T. A. Jacobson, S. Liberati, D. Mattingly, F.W. Stecker,
Phys.Rev. Lett. \textbf{93}, 021101 (2004); L. Maccione, Stefano Liberati,
Gunter Sigl, Phys. Rev. Lett. \textbf{105}, 021101 (2010); L. Shao, Z. Xiao,
Bo-Qiang Ma, Astropart. Phys. 33, 312 (2010); L. Maccione, S. Liberati, A.
Celotti, J. G. Kirk, P. Ubertini, Phys. Rev \ D\textbf{78},103003 (2008); M.
Galaverni, G. Sigl, Phys.Rev.D\textbf{78}, 063003 (2008); L. Maccione, S.
Liberati, J. Cosmol. Astropart. Phys. \textbf{0808}, (2008) 027; M. Galaverni,
G. Sigl, Phys. Rev. Lett. \textbf{100}, 021102 (2008); D.M. Mattingly, L.
Maccione, M. Galaverni, S. Liberati, G. Sigl, J. Cosmol. Astropart. Phys.
\textbf{1002}, (2010) 07; S.T. Scully, F.W. Stecker, Astropart. Phys.
\textbf{31}, 220 (2009).

\bibitem {Other}G. Gubitosi, G. Genovese, G. Amelino-Camelia, and A.
Melchiorri, Phys. Rev. D\textbf{ 82}, 024013 (2010); G. F. Giudice, M. Raidal,
A. Strumia, Phys.Lett.B \textbf{690}, 272 (2010); J. Alfaro, L. F. Urrutia,
Phys.Rev.D \textbf{81}, 025007 (2010); B. Withers, Class.Quant.Grav.
\textbf{26}, 225009 (2009); M. Visser, Phys.Rev.D \textbf{80}, 025011 (2009);
S. M. Carroll, Timothy R. Dulaney, M. I. Gresham, H. Tam, Phys. Rev.
D\textbf{79}, 065012 (2009).

\bibitem {Klink2}F.R. Klinkhamer and M. Risse, Phys. Rev. D \textbf{77},
016002 (2008); F.R. Klinkhamer and M. Risse, Phys. Rev. D \textbf{77}, 117901 (2008).

\bibitem {Klink3}F. R. Klinkhamer and M. Schreck, Phys. Rev. D \textbf{78},
085026 (2008).

\bibitem {Interac1}V. A. Kostelecky and A.G.M. Pickering, Phys. Rev. Lett.
\textbf{91}, 031801 (2003); B. Altschul, Phys.Rev. D \textbf{70, }056005 (2004).

\bibitem {Interac2}C.D. Carone, M. Sher, and M. \ Vanderhaeghen, Phys. Rev. D
\textbf{74}, 077901 (2006); B. Altschul, Phys. Rev. D \textbf{79}, 016004 (2009).

\bibitem {Interac3}M.A. Hohensee, R. Lehnert, D. F. Phillips, R. L. Walsworth,
Phys. Rev. D \textbf{80}, 036010(2009); M.A. Hohensee, R. Lehnert, D. F.
Phillips, R. L. Walsworth, Phys. Rev. Lett. \textbf{102}, 170402 (2009); B.
Altschul, Phys. Rev. D \textbf{80}, 091901(R) (2009).

\bibitem {Bocquet}J.-P. Bocquet \textit{et at.}, Phys. Rev. Lett.
\textbf{104,} 241601 (2010).

\bibitem {Tfinite2}R. Casana, M. M. Ferreira Jr., J. S. Rodrigues, Madson R.O.
Silva, Phys. Rev. D \textbf{80}, 085026 (2009).

\bibitem {Tfinite3}R. Casana, M. M. Ferreira, Jr, M. R.O. Silva, Phys.Rev. D
\textbf{81}, 105015 (2010).

\bibitem {Prop}R. Casana, M.M. Ferreira Jr, A. R. Gomes, P. R. D. Pinheiro,
Phys. Rev. D \textbf{80}, 125040 (2009).

\bibitem {Qasem}Q. Exirifard, "Cosmological birefringence constraints on
light", arxiv:1010.2054.

\bibitem {Altschul}B. Altschul, Phys. Rev. Lett. \textbf{98}, 041603 (2007).

\bibitem {Kob}A. Kobakhidze and B.H.J. McKellar, Phys. Rev. D \textbf{76},
093004 (2007); R. Casana, M.M. Ferreira Jr, Carlos E. H. Santos, Phys. Rev. D
\textbf{78}, 105014 (2008).

\bibitem {Sexl}R. U. Sexl and H.K. Urbantke, "Relativity, Groups, Particles:
special relativity and relativistic symmetry in field and particle physics",
Springer-Verlag, New York (1992).

\bibitem {Veltman}M. Veltman, "Quantum Theory of Gravitation", in Methods in
Field Theory, edited by R. Bailian and J. Zinn-Justin (North-Holland Publising
Company and World Scientific Publising Co Ltd, Singapore, 1981).
\end{thebibliography}
\end{document}